\documentclass[aps,prd,showpacs,floatfix,preprint]{revtex4}
\usepackage{amsmath,bm}
\usepackage{graphicx,url}
\usepackage{epsfig}
\usepackage{subfigure}
\usepackage{color}
\usepackage[normalem]{ulem}  

\begin{document}
\title{Role of nuclear physics in oscillations of magnetars}
\author{Rana Nandi$^{1}$, Prasanta Char$^{2}$, Debarati Chatterjee$^{3,}$, and 
Debades Bandyopadhyay$^{2}$}
\affiliation{$^{1}$Frankfurt Institute for Advanced Studies, Ruth Moufang 
Strasse 1, 60438 Frankfurt am Main, Germany}
\affiliation{$^{2}$Astroparticle Physics and Cosmology Division and Centre for 
Astroparticle Physics, Saha Institute of Nuclear Physics, HBNI, 
1/AF Bidhannagar, Kolkata, 700064, India}
\affiliation{$^{3}$Laboratoire de Physique Corpusculaire, ENSICAEN, 5 Boulevard Mar\'echal Juin,  F-14050 Caen, France}

\begin{abstract}
Strong magnetic fields have important effects on the crustal properties of magnetars.
Here we study the magneto-elastic oscillations of magnetars taking into consideration the effect of strong
magnetic fields on the crustal composition (magnetised crust).
We calculate global magneto-elastic (GME) modes as well as modes confined to 
the crust (CME) only. The ideal magnetohydrodynamics is 
adopted for the calculation of magneto-elastic oscillations of magnetars with
dipole magnetic fields. The perturbation equations obtained in general 
relativity using Cowling approximation are exploited here for the study of 
magneto-elastic oscillations. Furthermore, deformations due to magnetic fields 
and rotations are neglected in the construction of equilibrium models for
magnetars.   
The composition of the crust directly affects its shear modulus which we calculate using three
different nucleon-nucleon interactions:  SLy4, SkM and Sk272. The shear modulus of the crust is found to be
enhanced in strong magnetic fields $\geq 10^{17}$ G for all those Skyrme interactions. It is noted
that the shear modulus of the crust for the SLy4 interaction is much higher than those of the SkM and Sk272 interactions
in presence of magnetic fields or not.
Though we do not find any appreciable change 
in frequencies of fundamental GME and CME modes with and without magnetised crusts, frequencies of first overtones of 
CME modes are significantly affected in strong magnetic fields $\geq 10^{17}$ G. 
However, this feature is not observed in frequencies of first 
overtones of GME modes. 
As in earlier studies, it is also noted that the effects of crusts on 
frequencies of both types of maneto-elastic modes disappear when the
magnetic field reaches the critical field ($B > 4 \times 10^{15}$ G).  
Frequencies of GME and CME modes calculated with magnetised 
crusts based on all three nucleon-nucleon interactions, stellar models and
magnetic fields, are compared with frequencies of observed quasi-periodic 
oscillations (QPOS) in SGR 1806-20 and SGR1900+14. As in earlier studies, this comparison 
indicates that GME modes are essential to explain all the frequencies as CME modes can explain
only the higher frequencies.
\end{abstract}

\maketitle

\section{Introduction}
Soft gamma repeaters (SGRs) are characterised by their sporadic and short bursts
of soft gamma rays. Luminosities in these bursts could reach as high as 
$\sim 10^{41}$ ergs s$^{-1}$. There are about 14 SGRs known observationally 
\cite{kas}.   
Evidence of stronger emissions of gamma rays from SGRs was observed in 
several cases. These events are known as giant flares in which luminosities 
are $\sim 10^{44}-10^{46}$ ergs s$^{-1}$. So far three cases of giant flares 
were reported and those are SGR 0526-66, SGR 1900+14 and SGR 1806-20 
\cite{hurley99,barat03,israel05,watts11,watts07}. In giant flares, the 
early part of the spectrum 
was dominated by a hard flash of shorter duration followed by a softer decaying 
tail of a few hundreds of seconds. 

SGRs are very good candidates for magnetars which are neutron stars with very 
high surface magnetic fields $\sim 10^{15}$ G \cite{duncan92,Duncan98,kouv98}.
Giant flares might be caused by the evolving magnetic field and its stress on 
the crust of magnetars. It was argued that starquakes associated with 
giant flares could excite Global Seismic Oscillations (GSOs)\cite{Duncan98}. 
Torsional shear modes of magnetars with lower excitation energies would be 
easily 
excited. In this case, oscillations are restored by the Coulomb forces of 
crustal ions. Furthermore, the torsional shear modes have longer damping 
times.  
These findings implied that QPOs might be shear modes of magnetars 
\cite{Duncan98}. 
Quasi-periodic oscillations were found in the decaying tail of giant 
flares of SGR 1806-20 and SGR 1900+14. Detected frequencies for SGR 1806-20
are 18, 26, 29, 92.5, 150, 626.5 and 1837 Hz \cite{israel05,watts06,stroh06}, whereas
for SGR 1900+14 detected frequencies are 28, 53.5, 84 and 155 Hz \cite{watts05}. 
Huppenkothen et al \cite{Huppenkothen14a, Huppenkothen14b, Huppenkothen14c} recently analyzed short bursts
of some SGRs and found QPOs with frequencies 93 and 127 Hz in SGR J1550-5418 \cite{Huppenkothen14a} and with 57 Hz in
SGR 1806-20 \cite{Huppenkothen14b}.

It was noted from earlier theoretical models of QPOs that the observed 
frequencies, in particular higher frequencies, could be explained reasonably 
well using pure shear modes as well as CME modes 
\cite{watts11,Duncan98,piro05,sotani07,samuel07,sotani08a,sotani08b,steiner09}.
On the
other hand, lower frequencies of observed QPOs might be connected to Alfv\'{e}n 
modes of the fluid core. This makes the study of the oscillations of magnetar
crusts more difficult.  There were attempts to explain frequencies of QPOs using
Alfv\'{e}n oscillations of the fluid core without considering a crust 
\cite{sotani08a,Levin07,colaiuda09,cerd09}. Levin \cite{Levin06,Levin07} first pointed out that the
strong magnetic fields of magnetars should couple the Alfv\'{e}n oscillations of fluid core with
the oscillations in the solid crust. After that many authors studied 
the problem in detail \cite{levin11,levin12,colaiuda11,gabler12,gabler13a}.
The magnetohydrodynamics (MHD) coupling between the crust and core causes pure crustal modes to decay by emitting
Alfv\'{e}n waves into the core.
It was argued that CME modes might appear in GSOs and
explain frequencies of observed QPOs for not very strong magnetic fields 
despite all these complex problems \cite{Sotani11}. 
But global modes are expected to couple to the Alfv\'{e}n continuum in the core, and leads to the 
damping of the modes. Simulations to simplified models show 
\cite{Levin07,levin11} that CME oscillations are efficiently damped in the 
Alfv\'{e}n continuum 
as the crust reacts to the motion of the core. Consequently long-lived QPOs can
be generated at 
special points of the continuum (turning points and edges of continuum).

Several groups also 
studied the effect of neutron superfluidity and/or proton superconductivity of 
the crust and/or core on the calculated frequencies of magnetars 
\cite{Andersson09, Passamonti13, Gabler13b}. It was noted that neutron 
superfluidity enhanced fundamental frequencies of magneto-elastic oscillations.
On the other hand, it was argued that proton superconductivity could be 
destroyed in magnetic fields $> 5 \times 10^{16}$ G \cite{sinha}.

Nuclear physics of crusts plays an important role on the magneto-elastic modes
of magnetars. In particular, the effects of the nuclear symmetry energy
on the CME frequencies were investigated recently \cite{steiner09}. It 
may be worth noting here that CME mode frequencies are sensitive to the 
shear modulus of neutron star crusts. Furthermore, the shear modulus 
strongly depends on the composition of neutron star crusts. In earlier studies 
of magneto-elastic modes the effect of magnetic field on the 
composition of the crust was not considered. 
Surface magnetic fields as large as $\sim 10^{15}$ G have been reported in 
magnetars. 
Further, indirect estimates using the scalar virial theorem does not exclude
internal magnetic fields up to $10^{18}$.
Such large magnetic fields in magnetars may influence the ground state properties of neutron star 
crusts. Recently, we have investigated the influence of Landau quantisation of
electrons on the compositions and equations of state (EoS) of outer and inner 
crusts and obtained appreciable changes in those properties when the magnetic 
field is very strong ($B\geq 10^{17} G$) \cite{nandi11a,nandi11b}.
This, in turn, might influence the shear modulus of crusts and thereby magneto-elastic
frequencies of magnetars. This motivates us to study these mode
oscillations of magnetars using magnetised crusts. We define the crust to be
magnetised (non-magnetised)
when the effect of magnetic field on the crustal composition is considered 
(not considered).

We organise the paper in the following way. We describe models for 
calculating oscillation modes, shear modulus and compositions and EoS of 
magnetised crusts in Sec. II. Results of this calculation are discussed in Sec. III.
Section IV gives the summary and conclusions.

\section{Formalism}
QPOs were investigated in 
Newtonian gravity \cite{Duncan98,piro05,carrol86,mcder88} as well as general 
relativity \cite{sotani07,sotani08a,Sotani11,schumaker83,messios01} with and 
without magnetic fields. In many of
those calculations, the magnetised crust was decoupled from the fluid core.
But the magnetic field strongly couples the crust to the core and we need to 
calculate magneto-elastic modes.

Here we first study the effects of magnetised crusts on the magneto-elastic modes confined to the crust (CME) only, 
by considering a free slip between the crust and the core. Next,
we calculate the global magneto-elastic (GME) modes where coupling between the crust 
and the core has been considered. Mode frequencies are calculated following the 
model of Refs. \cite{Sotani06,sotani07,messios01}. The spherically 
symmetric general relativistic model of Sotani et al. \cite{Sotani06} adopted
in this calculation is a simplified one compared with the state-of-the-art 
general relativistic magnetohydrodynamical (MHD) model \cite {gabler12}. 
Furthermore, we do not consider the coupling to the Alfv\'{e}n continuum within 
the framework of this study, as the aim of this work primarily is to investigate
the influence of magnetised crusts on QPOs.

It is well known that a strong magnetic field breaks the spherical symmetry of 
a neutron star
due to anisotropy of the energy momentum tensor \cite{Bocquet95}. Hence the isotropic Tolman-Oppenheimer-Volkoff (TOV)
equations are no longer applicable for computing the mass-radius relations for 
polar magnetic fields $\sim 10^{17}$ G. 

Ideally for large magnetic fields, one must then calculate the neutron star structure using the anisotropic stress-energy tensor and solving equations
for hydrostatic equilibrium \cite{Chatterjee15}. 
Although this approximation is reasonable for magnetic fields 
$< 10^{17}$ G, the deviations from spherical symmetry
become non-negligible for higher fields.
However, the aim of this work is to study the relative changes in the 
mode frequencies due to magnetic fields. For this reason, we neglect the deformation of the neutron star and assume it to be 
spherically symmetric. 
The metric used to determine equilibrium stellar models has the form,
\begin{equation}
ds^2 = - e^{2\Phi} dt^2 + e^{2\Lambda} dr^2 + r^2 \left( d{\theta}^2 + 
sin^2{\theta} d{\phi}^2 \right)~.
\end{equation}
The equilibrium models are obtained by solving the Tolman-Oppenheimer-Volkoff
(TOV) equation with a perfect fluid EoS. 

Here we consider an axisymmetric poloidal magnetic field generated by four 
current $J_{\mu} = (0,0,0,J_{\Phi})$ and expand the four-potential 
into vector spherical harmonics as 
$A_{\mu} = a_{\ell_m}(r) sin{\theta} {\partial_{\theta}}P_{\ell_m}(cos{\theta})$.

The perturbed equations are obtained by 
linearising the equations of motion of the fluid and the magnetic induction 
equation \cite{sotani07,messios01}. Torsional modes are incompressible and do 
not 
result in any appreciable density perturbation in equilibrium stars. 
Consequently, one may adopt the relativistic Cowling approximation 
and neglect metric perturbations $\delta g_{\mu\nu}$=0 \cite{mcder83}. 
We consider axial type perturbation in the four velocity and
the relevant perturbed matter quantity 
is the $\phi$-component of the perturbed four velocity 
$\partial {u^{\phi}}$ \cite{sotani07}
\begin{equation}
\partial {u^{\phi}} = e^{-\Phi} \partial_t {\cal{Y}}(t,r) {\frac{1}{sin{\theta}}}{\partial_{\theta}}
P_l(cos{\theta})~,
\end{equation}
where $\partial_t$ and $\partial_{\theta}$ correspond to partial derivatives
with respect to time and $\theta$, respectively, $P_l(cos{\theta})$ is 
the Legendre polynomial of order $l$ and ${\cal{Y}}(t,r)$ is the angular 
displacement of the matter. It is to be noted that the radial and angular
variations of azimuthal displacement of stellar matter lead to
shears of the crystal lattice in neutron star crusts which are described by
the shear tensor $S_{\mu\nu}$ \cite{schumaker83}. Further, the shear stress tensor is 
given by $T_{\mu\nu} = - 2 {\mu} S_{\mu\nu}$, where $\mu$ is the isotropic 
shear modulus. The linearised equations of motion includes the contribution of 
$\delta T_{\mu\nu}$ \cite{sotani07}. 
 
Assuming a harmonic time dependence for ${\cal Y}(t,r) = e^{i\omega t} {\cal Y}
(r)$ and neglecting $\ell \pm 2$ terms, one obtains the eigenvalue equation for
the mode frequency \cite{sotani07}

\begin{multline}
 \left[\mu + (1 + 2 \lambda_1)\frac{{a_1}^2}{\pi r^4}\right]{\mathcal Y}''
    + \Bigg\{\left(\frac{4}{r} + \Phi' - \Lambda'\right)\mu \\
   + \mu' + (1 + 2\lambda_1)\frac{a_1}{\pi r^4}\left[\left(\Phi'
- \Lambda'\right)a_1
+ 2{a_1}'\right]\Bigg\}{\mathcal Y}' \\
    + \Bigg\{\left[\left(\epsilon + p +
(1 +2\lambda_1)\frac{{a_1}^2}{\pi r^4}
\right)e^{2\Lambda}
     - \frac{\lambda_1 {{a_1}'}^2}{2\pi r^2}\right]\omega^2 e^{-2\Phi}
 \\
    -(\lambda-2)\left(\frac{ \mu e^{2\Lambda}}{r^2}
- \frac{\lambda_1{{a_1}'}^2}{2\pi r^4}\right)\\
     + \frac{(2 + 5\lambda_1)a_1}{2\pi r^4}\left[\left(\Phi'
- \Lambda'\right){a_1}' + {a_1}''\right]
     \Bigg\}{\mathcal Y} = 0~, 
\label{eigen}
\end{multline}
where $\lambda = \ell (\ell + 1)$ and 
$\lambda_1 = - \ell (\ell + 1)/(2\ell - 1)(2\ell + 3)$.
Equation (\ref{eigen}) reduces to the non-magnetic case when we put $a_1=0$ 
\cite{sotani07}. Sotani et al. \cite{sotani08b} showed that the $\ell \pm 2$
truncation worked well for oscillations confined to the crust only. 
The eigenvalue equation for modes confined to the crust was solved using a two 
dimensional numerical method where $\ell \pm 2$ terms were not
truncated \cite{sotani08b}. It was demonstrated that results were unaffected 
whether $\ell \pm 2$ terms were truncated or not.  
With suitable choice of new variables, Eq.(\ref{eigen}) results in a system of
first order ordinary differential equations \cite{sotani07}. For 
magnerto-elastic
modes confined to the crust, we impose a zero traction
boundary condition at the interface between the core and the crust as well
as the zero torque condition at the surface \cite{sotani07,Sotani11}. These 
conditions imply ${\cal Y}' =0$ at the surface ($r=R$) of the star and the 
interface ($r=R_c$) of the crust and core. For the GME modes the 
boundary condition  at the surface is the same as CME modes 
\cite{sotani07,gabler12}.
The other boundary condition is the regularity at the center 
($\mathcal{Y}\sim r^{\ell-1}$). 
Finally, we estimate eigenfrequencies by solving two first order differential 
equations.

The knowledge of the shear modulus of magnetised crusts is an 
important input in the eigenvalue equation [Eq.(\ref{eigen})] for the 
CME mode calculation. Here we adopt the following expression of the shear 
modulus at zero temperature \cite{ogata90,stroh91} 
\begin{equation}
\mu = 0.1194 \frac{n_i (Ze)^2}{a}~,
\label{shr}
\end{equation}
where $a = [3/(4 \pi n_i)]^{1/3}$, $Z$ is the atomic number of a nucleus and 
$n_i$ is
the ion density. This form of the shear modulus was obtained by assuming a bcc
lattice and performing directional averages \cite{hansel01}. Further the 
dependence of the shear modulus on temperature was also investigated with the 
Monte Carlo sampling technique by Strohmayer et al. \cite{stroh91}. The 
composition and equation of state of neutron star crusts are essential 
ingredients 
for the calculation of the shear modulus as it is evident from Eq.(\ref{shr}). 

Now we describe the ground state properties in outer 
and inner crusts in the presence of strong magnetic fields. 
The outer crust is composed of nuclei immersed in a uniform background of a 
non-interacting electron gas. Neutrons start coming out of nuclei when the
neutron drip point is reached. This is the beginning of the inner crust where
nuclei are placed both in free neutrons as well as electrons. To minimize the 
Coulomb energy nuclei are 
arranged in a bcc lattice in neutron star crusts \cite{Bps71}. 

The ground state properties of matter of the inner crust is described using the
Thomas-Fermi model \cite{nandi11a}. The spherical cell that contains
neutrons and protons does not define a nucleus. We adopt the procedure of 
Bonche, Levit and Vautherin to subtract the free neutron gas of the cell 
and obtain the nucleus 
\cite{bonche84,bonche85,sil02}. 

For neutron star crusts in strongly quantising magnetic fields
we showed earlier that the Landau quantisation of electrons strongly
influenced ground state properties of neutron star crusts in strong magnetic
fields $\sim 10^{17}$ G \cite{nandi11a,nandi11b}. Energy and number densities 
of electrons are affected by the phase space modifications due to Landau 
quantisation of electrons. 
It is to be noted that protons are only influenced by 
magnetic fields through the charge neutrality condition.

\section{Results and Discussions}
We already investigated the composition and EoS of ground state matter in
neutron star crusts in strong magnetic fields \cite{nandi11a,nandi11b}. We 
noted that the electron number density in the outer crust was enhanced compared
with the field free case when a few Landau levels were populated for magnetic
fields $> 4.414 \times 10^{16}$ G \cite{nandi11a}. It was observed that this 
enhancement grew stronger when only the zeroth Landau level was populated at a
magnetic field strength of 4.414 $\times 10^{17}$ G. Consequently, we found
modifications in the sequence of equilibrium nuclei which was obtained by 
minimising the Gibbs free energy per nucleon. It was noted
that some new nuclei such as $^{88}_{38}$Sr and $^{128}_{46}$Pd appeared
and some nuclei such as $^{66}$Ni and $^{78}$Ni disappeared in a magnetic field
of $B=4.414 \times 10^{16}$ G \cite{nandi11a} when we compared this with the 
zero field case. It was further
observed that the neutron drip point was shifted to higher density in presence 
of a strong magnetic field with respect to the field free case 
\cite{nandi11a}.
We also performed the calculation of the inner crust using the SLy4 and SkM 
nucleon-nucleon interactions \cite{nandi11b}.
In this case too, we calculated the equilibrium nucleus at each density point.
Like the outer crust in strong magnetic fields, the electron number
density was enhanced due to the electron population in the zero Landau level for
magnetic fields $\geq 10^{17}$ G which , in turn, led to a large proton fraction
because of charge neutrality. For magnetic fields $>10^{17}$ G, equilibrium
nuclei with larger mass and atomic numbers were found to exist in the crust
\cite{nandi11b}. The free energy per nucleon of the nuclear 
system was reduced in magnetic fields compared with the corresponding case 
without a magnetic field. Furthermore, it was noted that higher symmetry 
energy in the sub-saturation regime for the SLy4 interaction resulted in nuclei
with larger mass and atomic numbers than those of the SkM interaction.

In this paper, we perform calculations of shear modulus and magneto-elastic mode
frequencies using the SLy4, SkM and Sk272 nucleon-nucleon interactions. 
Saturation nuclear matter properties of those interactions are listed in 
Table I. It is evident from the table that those nucleon-nucleon interactions 
differ in the symmetry energy and its slope coefficient from one interaction 
to the other. It is to be seen how the behaviour of the symmetry energy 
and its
slope coefficient in the sub-saturation density would impact the compositions
of magnetised crusts, its shear modulus and finally magneto-elastic modes. 

We calculate the shear modulus using Eq.(\ref{shr}) and the above mentioned 
models of magnetised crusts. Figure 1 displays the shear modulus as
a function of mass density for three nucleon-nucleon interactions of 
Table I with and without magnetic fields. Here we have shown results for 
$B_{*}= B/B_c=10^4$ where $B_c=4.414 \times 10^{13}$ G, where $B$ denotes the magnetic field strength at the pole.
When the field 
strength is $< 10^{17}$ G, the shear modulus does not show any 
appreciable change from that of the zero field because of large numbers of 
Landau levels are populated in this case. As the field strength is increased, 
less numbers of Landau levels are populated. For $B_{*} = 10^4$ i.e. 
$4.414 \times 10^{17}$ G, the shear modulus is enhanced due to the
population of all electrons in the zeroth Landau level. In all three cases, the 
shear modulus increases with mass density well before the crust-core interface.
{It is observed from Fig. 1 that the shear modulus is highest for the
SLy4 nucleon-nucleon interaction whereas it is the lowest for the Sk272 
interaction. This can be understood by noting that the symmetry energy at
sub-saturation densities is highest for the SLy4 interaction. In this density
regime, the symmetry energy decreases from its value at the saturation density
according to the slope coefficient ($L$). As the SLy4 interaction has the
lowest value of $L$ (see Table I), it has the highest value of the symmetry 
energy among all three nucleon-nucleon interactions. Higher symmetry energy 
leads to higher 
proton fraction and consequently higher electron fraction due to the charge 
neutrality. Therefore, higher symmetry energy implies higher shear modulus as is
evident from Eq. (\ref{shr}).   
The shear modulus and shear speed $v_s=(\mu/\rho)^{1/2}$ are extrapolated to the
zero value at the crust-core interface for magnetised as well as non-magnetised
crusts.
At densities close to the crust-core boundary nuclei can
take various non-spherical shapes collectively known as nuclear pasta \cite{Ravenhall83, Hashimoto84, Nandi16}. 
As the detailed nature of this pasta phase is not fully settled and there is no calculation 
of the shear modulus of this phase yet and as the shear modulus should vanish at the crust-core boundary, we extrapolate 
the shear modulus and shear speed $v_s=(\mu/\rho)^{1/2}$ to the
zero value at the crust-core interface for magnetised as well as non-magnetised
crusts. This approach is similar to that of Ref. \cite{Sotani11} where an arbitrary fit was used
so that the shear modulus smoothly decreases to zero at the crust-core interface, in the absence of magnetic fields.
We generate profiles of the shear modulus as a function of
radial distance in a neutron star for calculating frequencies of magneto-elastic
modes. The shear
modulus profiles along with the profiles of energy density and pressure
are obtained by solving the TOV equation. In this context, we construct the EoS 
of dense nuclear matter in strong magnetic fields in neutron star core using a 
relativistic mean field model with the GM1 parameter set as described in 
Ref.\cite{chakra97,sinha09,glend91}. This EoS of dense nuclear matter
is matched with the EoS of the crust and used in the TOV equation. 

\subsection{CME modes}
First we study the magneto-elastic modes confined to the crust only. We investigate} the dependence
of these mode frequencies on the compositions and the shear modulus of magnetised crusts. Earlier all 
calculations were performed using non-magnetic neutron star crusts. Here we
exploit models of non-magnetic as well as magnetic crusts which were already 
described in this section. We consider CME modes of a neutron star 
of mass 1.4 $M_{\odot}$. Frequencies of fundamental ($n=0$, $\ell=2$) CME modes are plotted with magnetic fields in Fig. 2 for all 
three nucleon-nucleon interactions. Here $n$ gives the number of radial 
nodes in the eigenfunction ${\cal Y}(r)$, in the crust. It is observed that 
in each case 
the frequency increases very slowly with magnetic field
for $B^* < 100$. But for $B^*>100$, the frequency increases linearly with 
magnetic fields. This behavior was also observed in earlier studies 
\cite{Sotani06,sotani07}.
The frequencies corresponding to the 
SLy4 interaction for $B^{*} < 100$, are almost two times higher than those 
of the SkM and Sk272 interactions. This is the direct consequence of the higher value of shear modulus for the SLy4 
interaction
than the other two interactions. However, there are no differences between our
results with and without magnetised crusts. This shows that the increase in shear modulus due to magnetic
field is too small to change the fundamental modes even for very high fields 
($\gtrsim 10^{17}$ G).

Figure 3 shows frequencies of CME modes corresponding to $n=0$ 
plotted as a function of $\ell$ values for a 1.4 $M_{\odot}$ neutron star, 
magnetic field $B_*=10^4$ and all three nucleon-nucleon interactions. 
Furthermore, we calculate frequencies using the non-magnetic as well as magnetic
crusts. In all cases frequency increases with higher $\ell$ values. For
higher values of $\ell$, frequencies with magnetic crusts are found to be 
slightly smaller than those of non-magnetic crusts, i.e. when the effect of 
magnetic field on the crustal composition is neglected. 
This is true for all three nucleon-nucleon interactions used in this 
calculation. The small decrease in frequencies in case of magnetised crusts
is due to increase of radius ($R$) of the star for $B_* = 10^4$ as is evident from
Table II because fundamental frequencies are inversely proportional to $R$.

We continue our investigation on frequencies of first overtones ($n=1$) of
CME modes in the presence of magnetic fields. Frequencies of
first overtones are shown as a function of $\ell$ values for a neutron star of 
1.4 $M_{\odot}$, magnetic field $B_*=10^4$ and all nucleon-nucleon interactions 
of Table I in Fig. 4. It is observed that the frequencies
obtained with magnetised crusts are significantly suppressed compared with
those of non-magnetised crusts for each nucleon-nucleon interaction and for 
all values of $\ell$. This is understood if we remember the fact that the radius
of a star is sensitive to the crustal EoS. Since strong magnetic fields ($\gtrsim 10^{17}$G)
change the composition as well as EoS of the crust, the stellar radius as well as crustal thickness also get 
affected. In Table \ref{t:radii}, we have shown the radius ($R$) and the ratio of the crust thickness ($\triangle$R) to the radius of
a neutron star for $B*=0$ and $B_*=10^4$, for all three nuclear interactions.
From the table we see that the value of $\triangle$R/R is larger for $B*=10^4$ than for $B=0$.
It was noted that the ratio of the 
crust thickness to the radius of a neutron star was inversely proportional to
the frequencies of overtones \cite{sotani07}.
Hence, it explains why overtone frequencies are smaller for magnetised crusts with $B*=10^4$, even though the shear moduli are
little larger for this case than that of $B_*=0$.
The effects of nucleon-nucleon interactions are 
evident from the figure where the results of the SkM lie at the top 
and those of the SLy4 are at the bottom. 
This can also be understood from Table \ref{t:radii} if we note that the ratio ($\triangle$R/R) 
is the highest for the SLy4 interaction and lowest for the SkM  interaction.

The dependence of frequencies of the fundamental mode and higher harmonics
on neutron star masses is demonstrated in Fig. 5 for the SLy4, SkM and Sk272
nucleon-nucleon interactions. Here the frequencies 
corresponding to $n=0$ and $\ell = 2,3,4$ are shown as a function of neutron 
star masses for a magnetic field $B=8 \times 10^{14}$ G. For all cases, 
frequencies of CME modes decrease with increasing mass, whereas
higher $\ell$ values lead to higher frequencies. It is observed from Fig. 5
that frequencies corresponding 
to (non)magnetic crusts based on the SLy4 nucleon-nucleon interaction are much
higher than those of other two nucleon-nucleon interactions. When the 
calculated frequencies are compared with the frequencies of observed QPOs, the 
latter might put a strong constraint on the EoS if masses of neutron 
stars are known accurately.  

Next, we compare the calculated frequencies of CME modes with
frequencies of observed QPOs. These comparisons are shown in 
Tables \ref{t:ft1} and \ref{t:ft2}. Here we have also included QPO of 57 Hz found
recently by Huppenkothen {\it et al.} \cite{Huppenkothen14b} in the short bursts of SGR 1806-20.
 For SGR1806-20,
our results in Table \ref{t:ft1}  are obtained using the magnetised crusts of 1.3, 1.4 
and 1.7
$M_{\odot}$ neutron stars based on the SLy4, SkM and Sk272 nucleon-nucleon 
interactions, respectively, and magnetic field $B=8 \times 10^{14}$ G. It is
noted that calculated frequencies below 93 Hz for each nucleon-nucleon 
interaction can not explain the observed frequencies whereas our results above
93 Hz are in very good agreement with observed QPO frequencies
\cite{israel05,stroh06,watts06}. Similarly for SGR 1900+14, 
we calculate CME mode frequencies using magnetised crusts of 1.7, 
1.2 and 1.2 $M_{\odot}$ neutron stars corresponding to the SLy4, SkM and Sk272 
nucleon-nucleon interactions
and $B= 4 \times 10^{14}$ G. These results are shown in Table \ref{t:ft2}. Our 
calculated frequencies for all three nucleon-nucleon interactions are 
in agreement with the observed QPO frequencies of SGR 1900+14
\cite{watts05}. 

\subsection{GME modes}
First we calculate pure Alfv\'{e}n modes of a neutron star of mass 
$1.4M_{\odot}$,
by ignoring the presence of the solid crust. In Fig. 6, we show the
pure Alfv\'{e}n mode corresponding to $n=0;\, \ell=2$ as a function 
of magnetic field ($B_*$). Here, $n$ stands for the number of radial nodes 
in the eigenfunctions, in the liquid core. We see that the frequency of this mode increases 
linearly with magnetic field and become equal to those of the CME modes above 
$B_{*}=100$. Next, we calculate corresponding GME mode 
frequencies for various magnetic fields, taking magnetic crusts into 
consideration. Magnetised crusts used here are calculated with the SLy4, 
SkM and Sk272 nucleon-nucleon interactions.
It is observed that at low magnetic fields global mode frequencies 
have higher values as compared to those of pure Alfv\'{e}n modes. The GME
modes are found to be confined to the core for low magnetic field strengths. 
This scenario is similar to the reflection of GME modes at the crust-core
interface as manifested in the
state-of-the-art model of Gabler et al. \cite{gabler12}. Consequently, this 
leads to 
higher frequencies for GME modes compared with those of pure Alfv\'{e}n modes. 
But at higher 
magnetic fields, GME mode frequencies merge with that of pure Alfv\'{e}n 
modes. This happens because at higher values of fields ($B \geq 4.14 \times 
10^{15}$) shear modulus
becomes negligible as compared to the magnetic effect ($\mu \ll B^2$); in other
words Alfv\'{e}n velocity ($B/\sqrt{4\pi\rho}$) becomes much larger compared to
the shear velocity ($\sqrt{\mu/\rho}$).  We also show the frequencies of 
CME modes for comparison. It is also evident from Fig. 6 that 
the effects of crusts on frequencies disappear at very high magnetic fields 
$B_{*}>100$ and oscillations become magnetic oscillations \cite{sotam}.    

To see the effects of magnetic crusts on GME modes we calculate 
these modes with and without magnetic crusts based on the SLy4, SkM and Sk272 
nucleon-nucleon interactions. Figures 7 and 8 show results for
modes with $n=0$ and $n=1$, respectively as a function of $\ell$ for a neutron 
star of mass $1.4M_{\odot}$ and magnetic field $B_*=10^4$. We can
see there is no significant change in frequencies if the crust is considered to
be magnetic. For fundamental modes in Fig. 7, there is no 
appreciable change in frequencies with and without magnetic crusts. Unlike
Fig. 3 for CME modes, GME modes are insensitive to the small change in R. 
In case of first overtones in 
Fig. 8, we do not find any appreciable effects of crusts on frequencies because
the magnetic field $B_{*}=10^4$ is so high that oscillations become magnetic 
oscillations.   

We also attempt to match the observed frequencies with those of calculated 
GME modes. The results are shown in Tables V and VI. 
For SGR1806-20, we compute frequencies using the magnetised crusts of 1.5, 
1.4 and 1.4 $M_{\odot}$ neutron stars based on the SLy4, SkM and Sk272 
nucleon-nucleon interactions, respectively, and magnetic field 
$B=3.1 \times 10^{15}$ G. These results are given by Table V. 
We find that the
calculated frequencies agree well with the lower and higher frequencies 
of observed QPOs. However, it is noted that large values of $n$ are needed to 
fit 
higher frequencies. This feature for higher frequencies was also obtained by
Sotani et al. \cite{Sotani06}. On the other 
hand, we exploit magnetised crusts of 1.4, 1.3 and 1.3 M$_{\odot}$ neutron
stars corresponding to the SLy4, SkM and Sk272 nucleon-nucleon interactions
and magnetic field $B=1.34\times 10^{15}$ G for SGR 1900+14. The Table VI
demonstrates that the calculated frequencies are in good agreement with the
observed frequencies of SGR 1900+14. We do not find any appreciable effects
of nucleon-nucleon interactions in either table. 

\section{Summary and Conclusions}
We have estimated frequencies of  global magneto-elastic modes as well 
as magneto-elastic modes confined to the crust only of magnetars assuming a dipole magnetic field configuration.  
Frequencies are computed using our models of magnetised crusts based on the
SLy4, SkM and Sk272 nucleon-nucleon interactions. Though the formalism
used in Sotani et al. \cite{Sotani06} and in this calculation are same, 
magnetised crusts are employed for the first time here. The shear modulus of 
magnetised crusts is found to be enhanced in strong magnetic fields 
$\sim 4.414 \times 10^{17}$ G because electrons populate the zeroth Landau 
level. It is observed that frequencies of the fundamental ($n=0$, $\ell=2$) 
CME mode are not sensitive to this enhancement in the shear modulus in strong 
magnetic fields. 
On the other hand, frequencies of first overtones 
($n=1$) of CME modes in the presence of strongly quantising magnetic 
fields are distinctly different from those of the field free case. 
It is shown that that this is related to the the ratio of the crust 
thickness to the radius of a magnetar. 
We have found that at $B_*=10^4$, the $\triangle R/R$ is increased by
$2-4\%$, which causes frequencies of overtones to decrease by $5-7\%$, for the
models we used here.
For {GME} modes, the effects of crusts disappear above a critical field
($B > 4 \times 10^{15}$ G) and oscillations become magnetic oscillations.
We have compared frequencies of CME and GME modes calculated using different stellar models, magnetic field strengths and 
magnetised crusts based
on three nucleon-nucleon interactions with frequencies of observed QPOs and 
conclude that the agreement is reasonable for SGR 1900+14 in both cases. 
However,
the calculated frequencies of CME modes do not match with lower
frequencies of SGR 1806-20, but can explain higher frequencies well. In the case of
GME modes, we find the opposite trends in fitting the frequencies
of SGR 1806-20. Finally new results that we have obtained would
be reproduced even in a sophisticated MHD calculation.


\newpage
\begin{table}[h]
\caption{Saturation nuclear matter properties of different Skyrme 
nucleon-nucleon interactions
used in this work such as saturation density ($\rho_0$), binding energy (BE),  
incompressibility (K), symmetry energy (J) and its slope coefficient (L)} 
\begin{center}
 \begin{tabular}{cccccc}
 \hline
 Parameter set& $\rho_0$& BE & K & J & L\\
 & (fm$^{-3}$) & (MeV)& (MeV) & (MeV) & (MeV) \\
 \hline
 SLy4 & 0.16 & 15.97 & 229.91 & 32.00 & 45.94\\
 SkM& 0.16 & 15.77 & 216.61 & 30.75 & 49.34\\
 Sk272& 0.155 & 16.28 & 271.51 & 37.40 & 91.67\\
 \hline
 \end{tabular}
 \label{t:param}
\end{center}
\end{table}
\begin{table}[h]
\begin{center}
\caption{ Radius and crust thickness for all three interactions at $B*=0$ and $B*=10^4$}
\begin{tabular}{ccc||cc}
\hline
   &\multicolumn{2}{c}{$B=0$}&\multicolumn{2}{c}{$B_*=10^4$}\\
 Set\quad\quad  &R(km)&\quad$\triangle$R/R\quad\quad&\quad\quad R(km)&\quad$\triangle$R/R\\
\hline 
 SLy4 & 13.972 & 0.096 & 13.987 & 0.100 \\
 SkM  & 13.875 & 0.086 & 13.892 & 0.088 \\
 Sk272& 13.910 & 0.089 & 13.927 & 0.092 \\
\hline               
\end{tabular}
\label{t:radii}
\end{center}
\end{table}
\newpage
\begin{table}[]
\begin{center}
\caption{Frequencies of CME modes calculated using magnetised 
crusts based on the SLy4, SkM and Sk272 nucleon-nucleon interactions 
are compared with observed QPO frequencies of SGR 1806-20 \cite{israel05,watts06,stroh06,Huppenkothen14b}. 
The magnetic field used in this calculation is $B = 8\times 10^{14}$ G. Here $f$, $n$ and $\ell$ 
represent frequency, radial node and angular node, respectively}
\begin{tabular}{cccccccccccc}
\hline
Observed & \multicolumn{11}{c}{Calculated frequency (Hz)}\\
frequency (Hz)&\multicolumn{3}{c}{SLy4}&&\multicolumn{3}{c}{SkM}&&\multicolumn{3}{c}{Sk272}\\
              &\multicolumn{3}{c}{$(1.3 M_{\odot}$)}&&\multicolumn{3}{c}{($1.4 M_{\odot}$)}&&\multicolumn{3}{c}{($1.7 M_{\odot}$)}\\
\hline 
             &  $f$  & n & $\ell$ &&   $f$  & n & $\ell$ &&   $f$  & n & $\ell$\\
        18   &  20.0 & 0 &    2   &&  13.0  & 0 &  2     &&  18.0  & 0 & 3 \\ 
        26   &   -   & - &    -   &&  20.7  & 0 &  3     &&  24.3  & 0 & 4 \\
        30   &  31.7 & 0 &    3   &&  27.8  & 0 &  4     &&  30.3  & 0 & 5 \\ 
        57   &  53.1 & 0 &    5   &&  55.0  & 0 &  8     &&  59.5  & 0 & 10\\
        92.5 &  94.1 & 0 &    9   &&  94.5  & 0 & 14     &&  93.9  & 0 & 16\\
       150   & 154.6 & 0 &   15   && 152.6  & 0 & 23     && 150.0  & 0 & 26\\
       626   & 627.9 & 1 &   16   && 626.9  & 1 & 27     && 625.9  & 1 & 34\\
      1838   &1834.5 & 4 &    2   &&1836.3  & 4 &  2     &&1841.9  & 4 & 2\\ 
\hline               
\end{tabular}
  \label{t:ft1}
  \end{center}
\end{table}
\newpage
\begin{table}[T]
\begin{center}
\caption{ Same as Table \ref{t:ft1} but for SGR 1900+14 \cite{watts05}. The magnetic field
used in this calculation is $B = 4\times 10^{14}$ G.}
\begin{tabular}{cccccccccccc}
\hline
Observed & \multicolumn{11}{c}{Calculated frequency (Hz)}\\
frequency(Hz)&\multicolumn{3}{c}{SLy4}&&\multicolumn{3}{c}{SkM}&&\multicolumn{3}{c}{Sk272}\\
             &\multicolumn{3}{c}{$(1.7 M_{\odot}$)}&&\multicolumn{3}{c}{($1.2 M_{\odot}$)}&&\multicolumn{3}{c}{($1.2 M_{\odot}$)}\\

\hline 
             &  $f$  & n & $\ell$ &&   $f$  & n & $\ell$ &&   $f$  & n & $\ell$\\
        28   &  28.4 & 0 &    3   &&  28.3  & 0 &  4     &&  26.6  & 0 & 4 \\ 
        54   &  56.7 & 0 &    6   &&  55.8  & 0 &  8     &&  52.5  & 0 & 8 \\
        84   &  84.2 & 0 &    9   &&  82.7  & 0 & 12     &&  83.8  & 0 &13 \\ 
       155   & 156.4 & 0 &   17   && 155.2  & 0 & 23     && 157.1  & 0 &25\\
\hline               
\end{tabular}
  \label{t:ft2}
  \end{center}
\end{table}
\begin{table}[]
\begin{center}
\caption{GME mode frequencies obtained using the magnetised 
crusts based on the SLy4, SkM and Sk272 nucleon-nucleon interactions 
are compared with observed frequencies in SGR 1806-20. The magnetic field used 
in this calculation is $B = 3.1 \times 10^{15}$ G.}
\begin{tabular}{cccccccccccc}
\hline
Observed & \multicolumn{11}{c}{Calculated frequency (Hz)}\\
frequency(Hz)&\multicolumn{3}{c}{SLy4}&&\multicolumn{3}{c}{SkM}&&\multicolumn{3}{c}{Sk272}\\
             &\multicolumn{3}{c}{$(1.5 M_{\odot}$)}&&\multicolumn{3}{c}{($1.4 M_{\odot}$)}&&\multicolumn{3}{c}{($1.4 M_{\odot}$)}\\

\hline 
             &  $f$  & n & $\ell$ &&   $f$  & n & $\ell$ &&   $f$  & n & $\ell$\\
        18   &  17.8 & 0 &    3   &&  18.2  & 0 &  3     &&  18.1  & 0 & 3 \\ 
        26   &  26.1 & 0 &    6   &&  26.1  & 0 &  6     &&  25.8  & 0 & 6 \\
        30   &  30.7 & 0 &    8   &&  30.7  & 0 &  8     &&  30.4  & 0 & 8 \\ 
        57   &  57.8 & 1 &    7   &&  57.1  & 1 &  6     &&  56.7  & 1 & 6\\
        92.5 &  93.0 & 4 &    2   &&  91.6  & 2 &  8     &&  94.5  & 2 & 9\\
       150   & 150.0 & 6 &    4   && 150.9  & 4 & 10     && 150.3  & 6 & 3\\
       626   & 624.3 &30 &    6   && 626.4  &28 &  6     && 628.6  &27 & 9\\
      1838   &1837.3 &96 &    5   &&1836.4  &97 &  2     &&1835.8  &87 & 10\\ 
\hline               
\end{tabular}
  \label{t:fg1}
  \end{center}
\end{table}
\newpage
\begin{table}[T]
\begin{center}
\caption{Same as Table \ref{t:fg1} but for the SGR 1900+14. The magnetic field adopted
here is $B= 1.34 \times 10^{15}$ G.}
\begin{tabular}{cccccccccccc}
\hline
Observed & \multicolumn{11}{c}{Calculated frequency (Hz)}\\
frequency(Hz)&\multicolumn{3}{c}{SLy4}&&\multicolumn{3}{c}{SkM}&&\multicolumn{3}{c}{Sk272}\\
             &\multicolumn{3}{c}{$(1.4 M_{\odot}$)}&&\multicolumn{3}{c}{($1.3 M_{\odot}$)}&&\multicolumn{3}{c}{($1.3 M_{\odot}$)}\\

\hline 
             &  $f$  & n & $\ell$ &&   $f$  & n & $\ell$ &&   $f$  & n & $\ell$\\
        28   &  28.0 & 1 &    6   &&  28.4  & 1 &  6     &&  28.0  & 1 & 5 \\ 
        54   &  54.7 & 3 &    8   &&  54.7  & 3 &  7     &&  53.6  & 2 &11 \\
        84   &  84.7 & 7 &    5   &&  84.1  & 6 &  6     &&  84.4  & 5 & 8 \\ 
       155   & 155.6 & 16 &    3   && 154.7  & 14 &  4     && 154.5  &11 & 9\\
\hline               
\end{tabular}
\label{t:fg2}
\end{center}
\end{table}
\newpage

\vspace{-5cm}

{\centerline{
\epsfxsize=12cm
\epsfysize=14cm
\epsffile{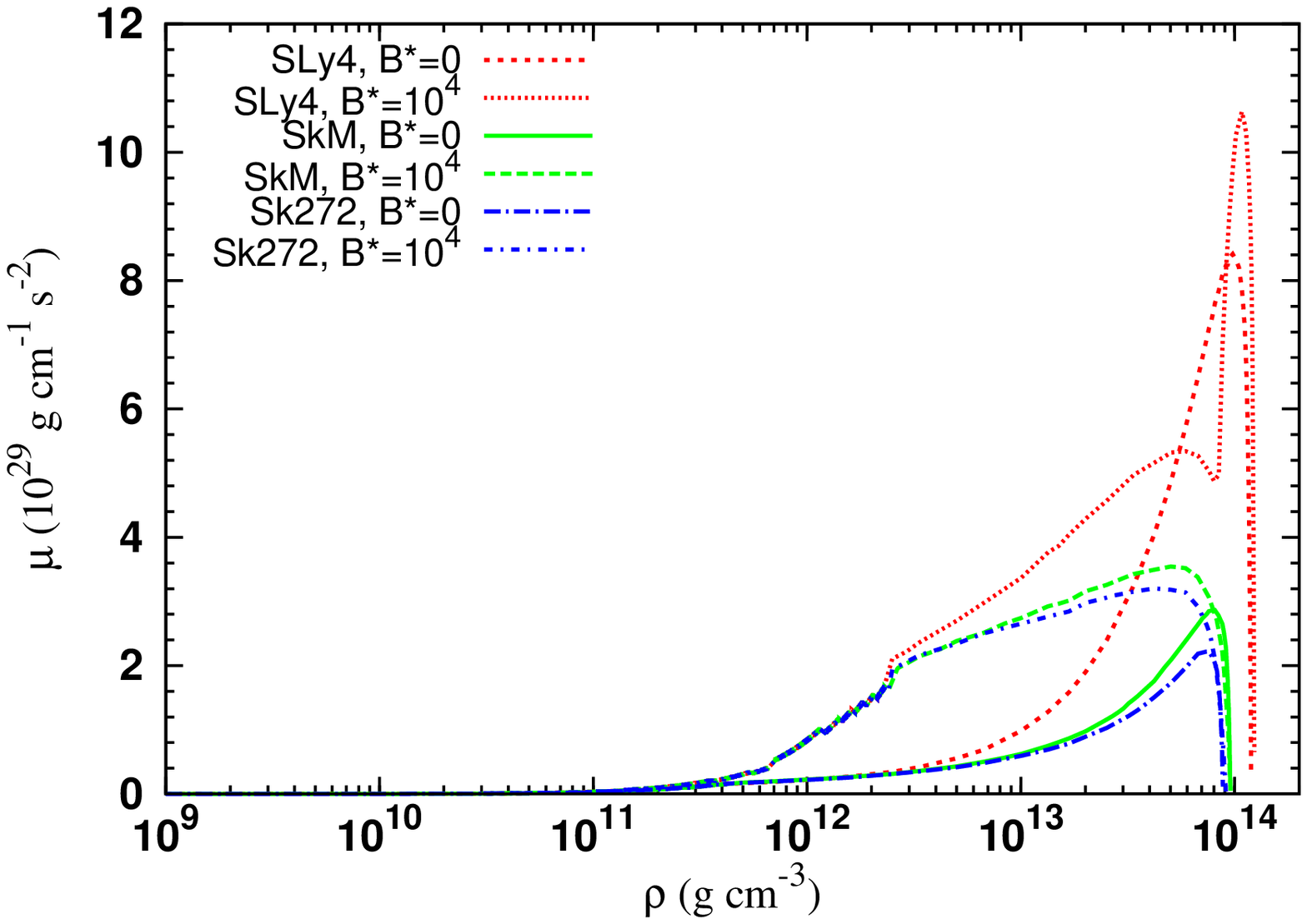}
\label{shrmdls}
}}

\vspace{4.0cm}

\noindent{\small{
FIG. 1.  
Shear modulus as a function of mass density for a neutron star of 
1.4 $M_{\odot}$ with magnetic fields $B_{*}=0$ and $B_{*}=10^4$ and 
Skyrme nucleon-nucleon interactions of Table I.
}}
  
\newpage

\vspace{-5cm}

{\centerline{
\epsfxsize=12cm
\epsfysize=14cm
\epsffile{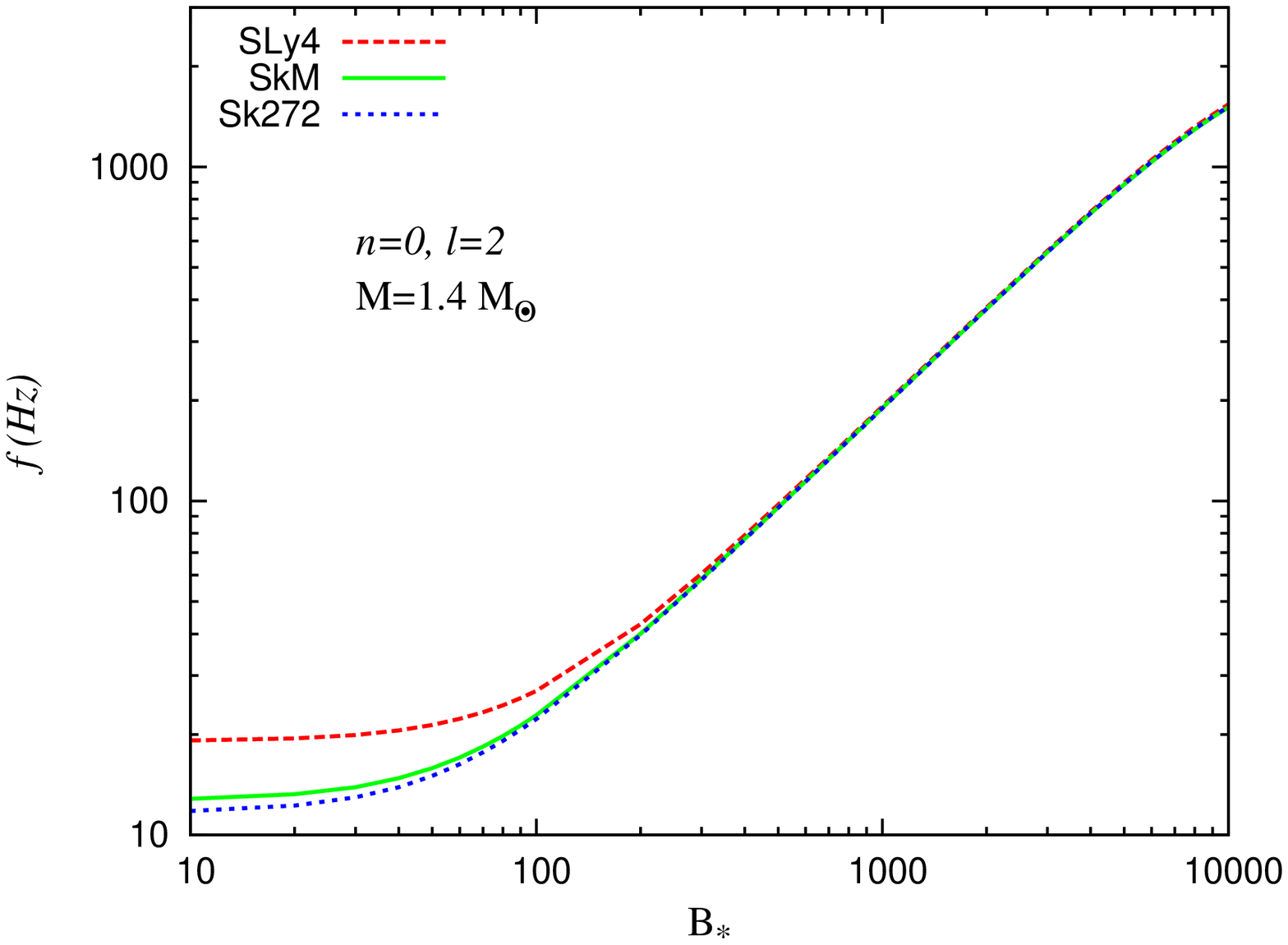}
}

\vspace{4.0cm}

\noindent{\small{
FIG. 2.  Frequency of fundamental ($n=0$, $\ell=2$) CME 
mode for  a neutron star of 1.4 $M_{\odot}$ is shown as a function of 
magnetic field $B_{*}=B/B_c$ where $B_{c}=4.414 \times 10^{13}$ G. Results of 
our calculations using the SLy4, SkM  and Sk272 nucleon-nucleon interactions 
are shown here.}}
\label{f:freq}
\newpage

\vspace{-5cm}

{\centerline{
\epsfxsize=12cm
\epsfysize=14cm
\epsffile{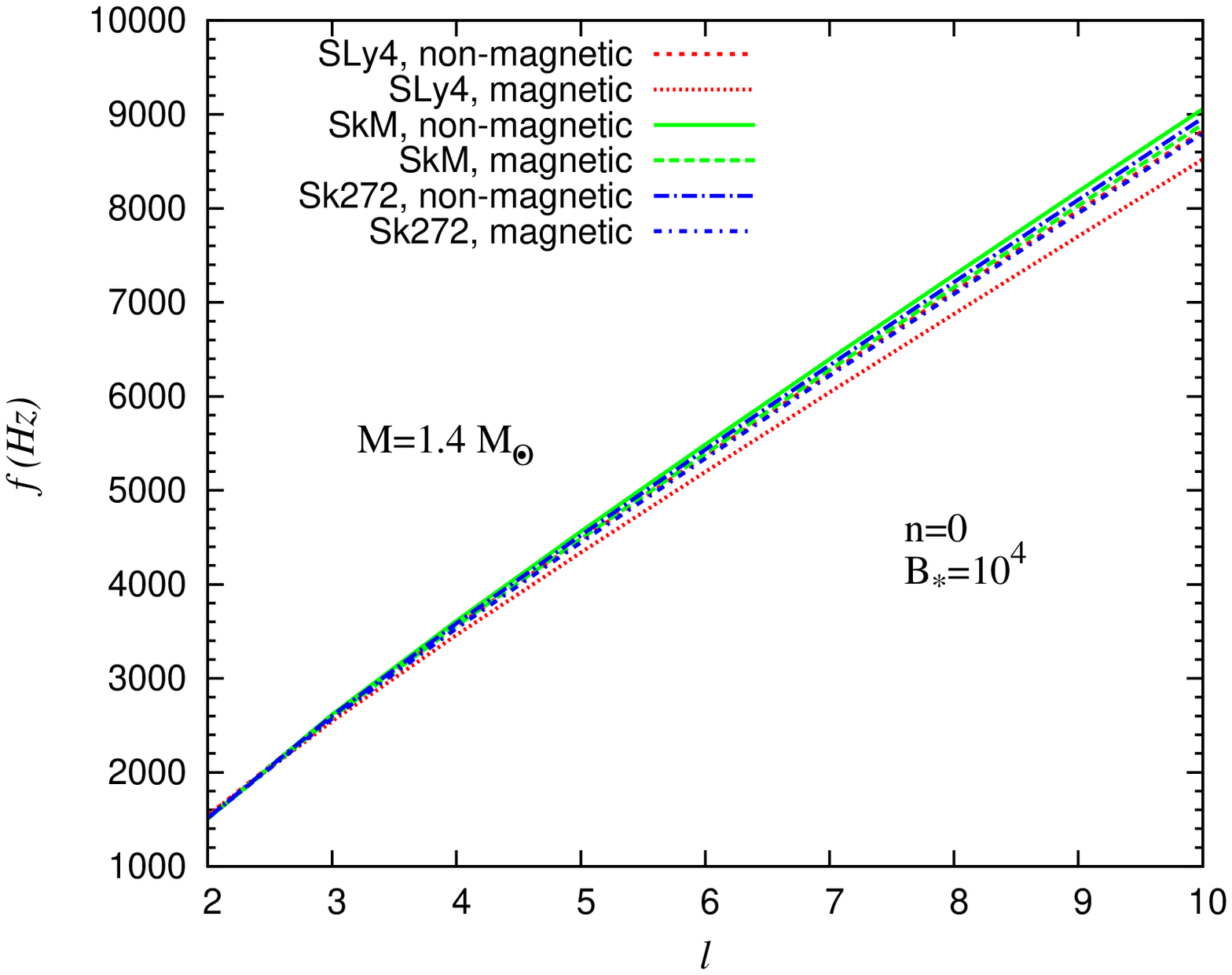}
}}

\vspace{4.0cm}

\noindent{\small{
Fig. 3. Fundamental frequencies ($n=0$) of CME modes are plotted 
as a function
of $\ell$ values with and without magnetic crusts of a 1.4 $M_{\odot}$ neutron 
star based on the SLy4, SkM and Sk272 nucleon-nucleon interactions  
for $B_{*}=10^4$.}}
\label{f:fvslbs0}
\newpage

\vspace{-5cm}

{\centerline{
\epsfxsize=12cm
\epsfysize=14cm
\epsffile{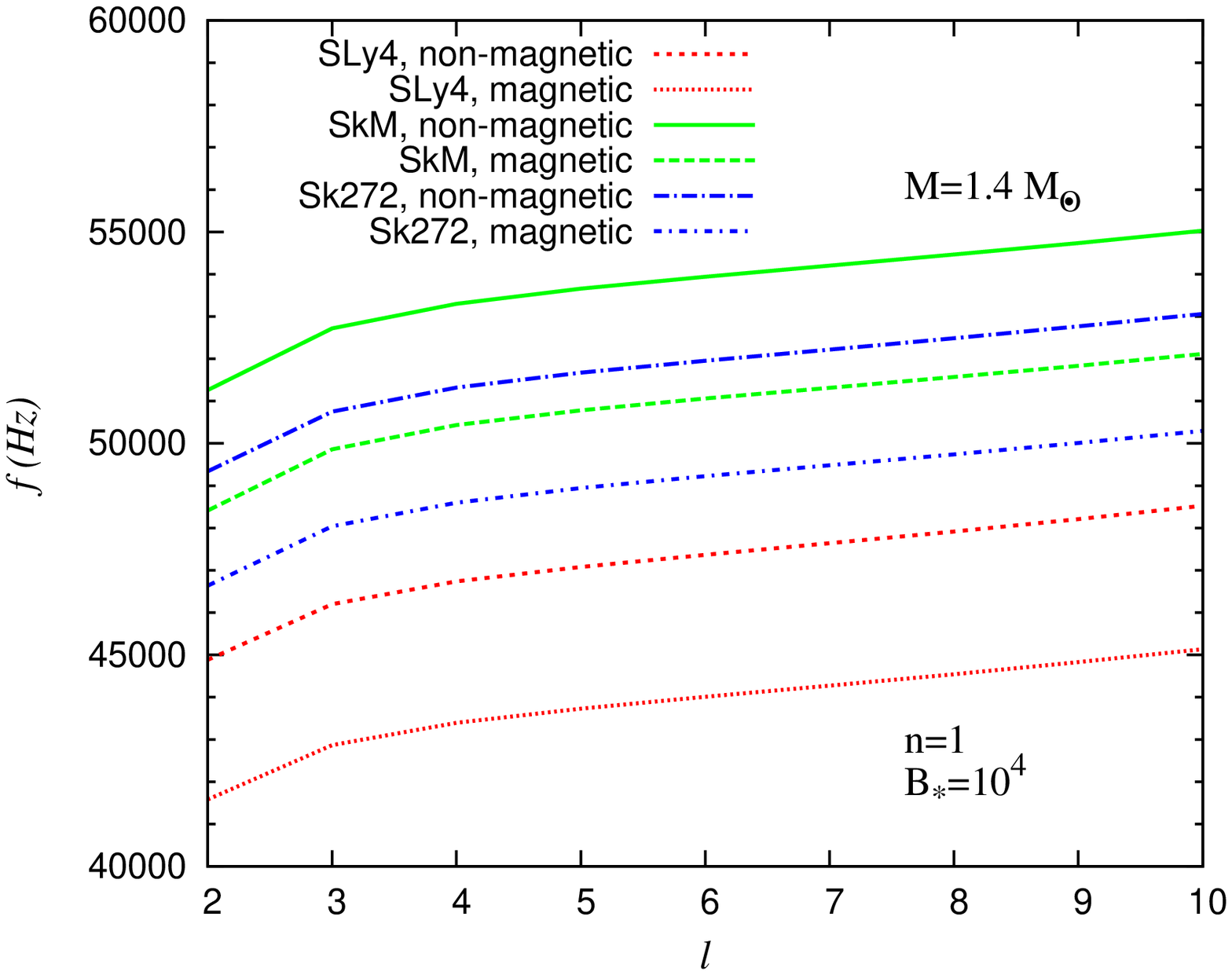}
}}

\vspace{4.0cm}

\noindent{\small{Fig. 4.
Frequencies of first overtones ($n=1$) of CME modes are shown 
as a function of $\ell$ values 
with and without magnetic crusts of a 1.4 $M_{\odot}$ neutron 
star based on the SLy4, SkM and Sk272 nucleon-nucleon interactions  
for $B_{*}=10^4$.}}
\label{f:fvslbs1}

\newpage

\vspace{-5cm}

{\centerline{
\epsfxsize=12cm
\epsfysize=14cm
\epsffile{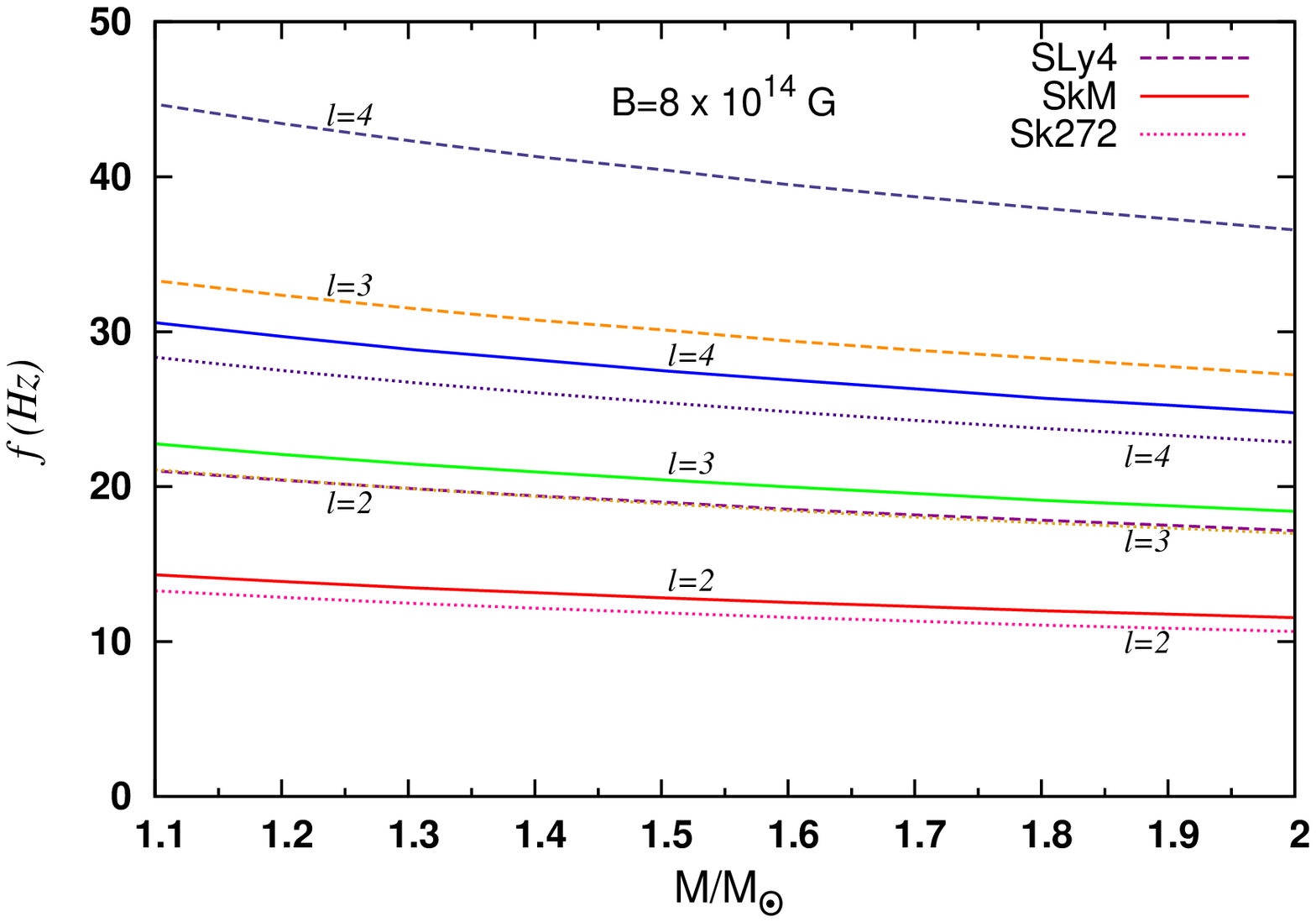}
}}

\vspace{4.0cm}

\noindent{\small{Fig. 5. Frequencies of CME modes corresponding to 
$n=0$ and $\ell=2,3,4$ are plotted as a function of neutron star mass for a 
magnetic field 
$B = 8 \times 10^{14}$ G using magnetised crusts based on the SLy4, SkM 
and Sk272 nucleon-nucleon interactions.}}
\label{f:fvsm}

\newpage

\vspace{-5cm}

{\centerline{
\epsfxsize=12cm
\epsfysize=14cm
\epsffile{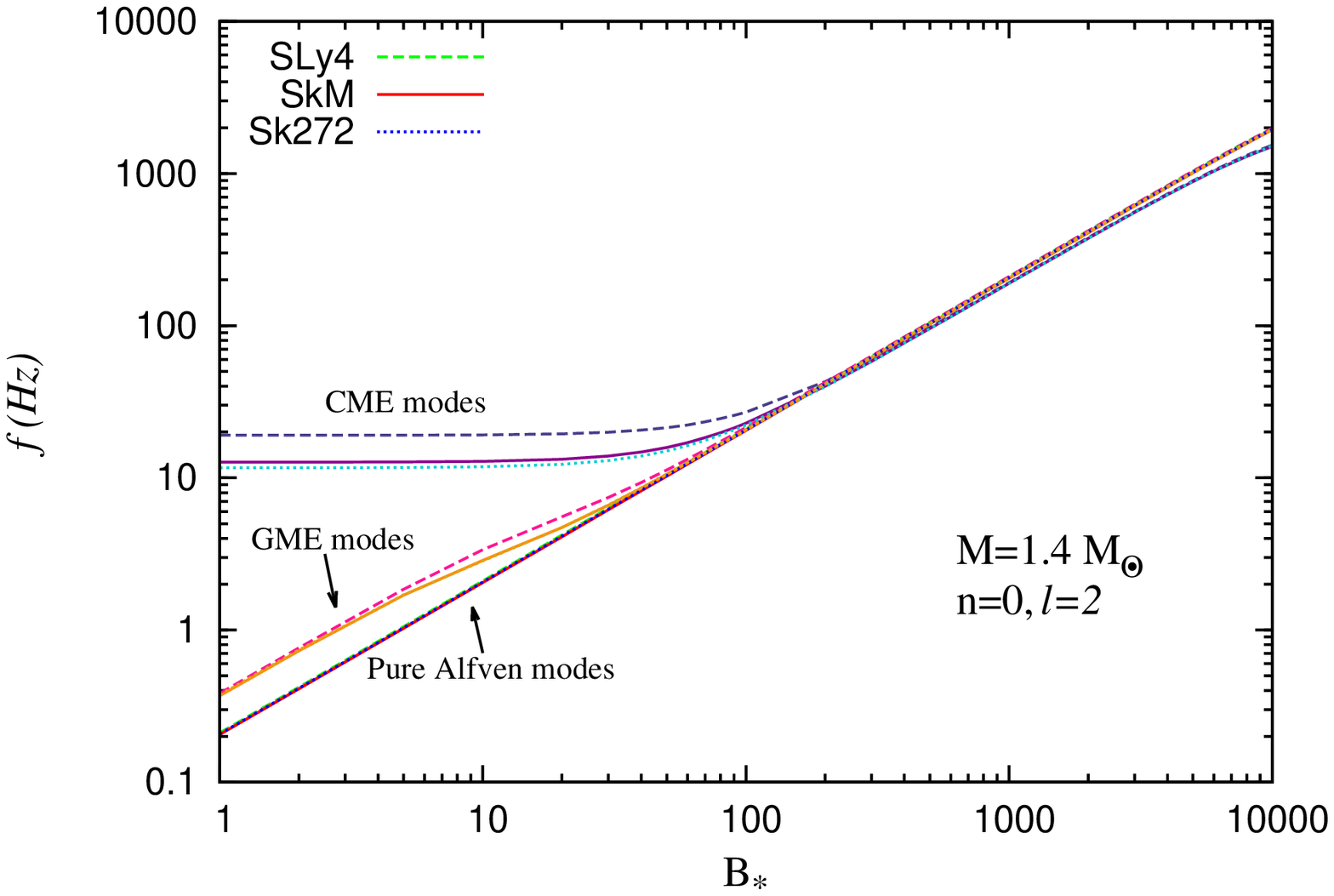}
}}

\vspace{4.0cm}

\noindent{\small{Fig. 6. 
Comparison of the GME frequencies with pure Alfv\'{e}n 
frequencies as well as CME frequencies is shown as a function
of magnetic field 
using the magnetised crusts based on the SLy4, SkM and
Sk272 nucleon-nucleon interactions.}}
\label{f:fcomp}
\newpage

\vspace{-5cm}

{\centerline{
\epsfxsize=12cm
\epsfysize=14cm
\epsffile{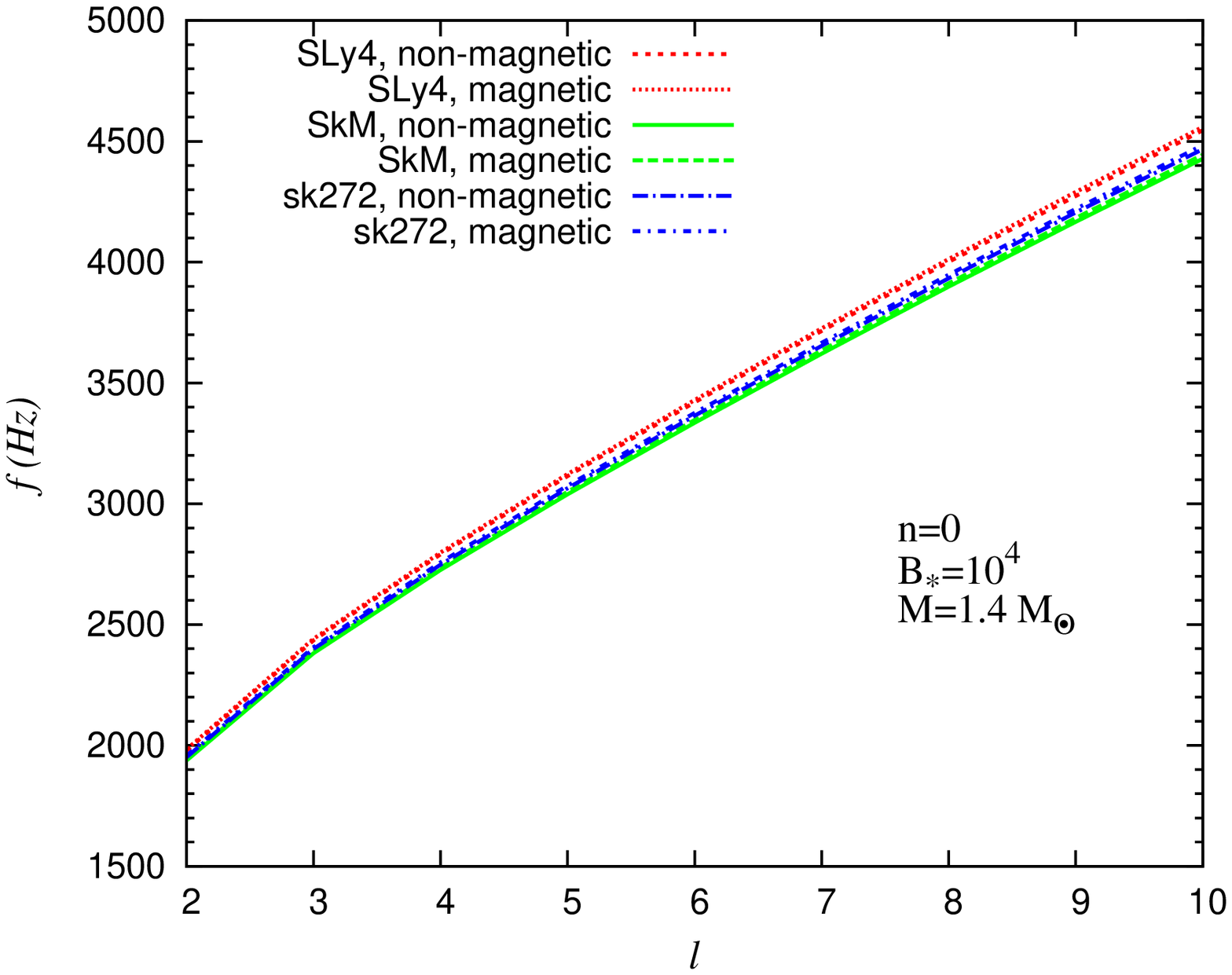}
}}

\vspace{4.0cm}

\noindent{\small{Fig. 7. 
GME mode frequencies for $n=0$ are shown as a function of $\ell$ 
values 
with and without magnetic crusts of a neutron star of mass $1.4M_{\odot}$
based on the SLy4, SkM and Sk272 nucleon-nucleon interactions for 
$B_*=10^4$.}}

\newpage

\vspace{-5cm}

{\centerline{
\epsfxsize=12cm
\epsfysize=14cm
\epsffile{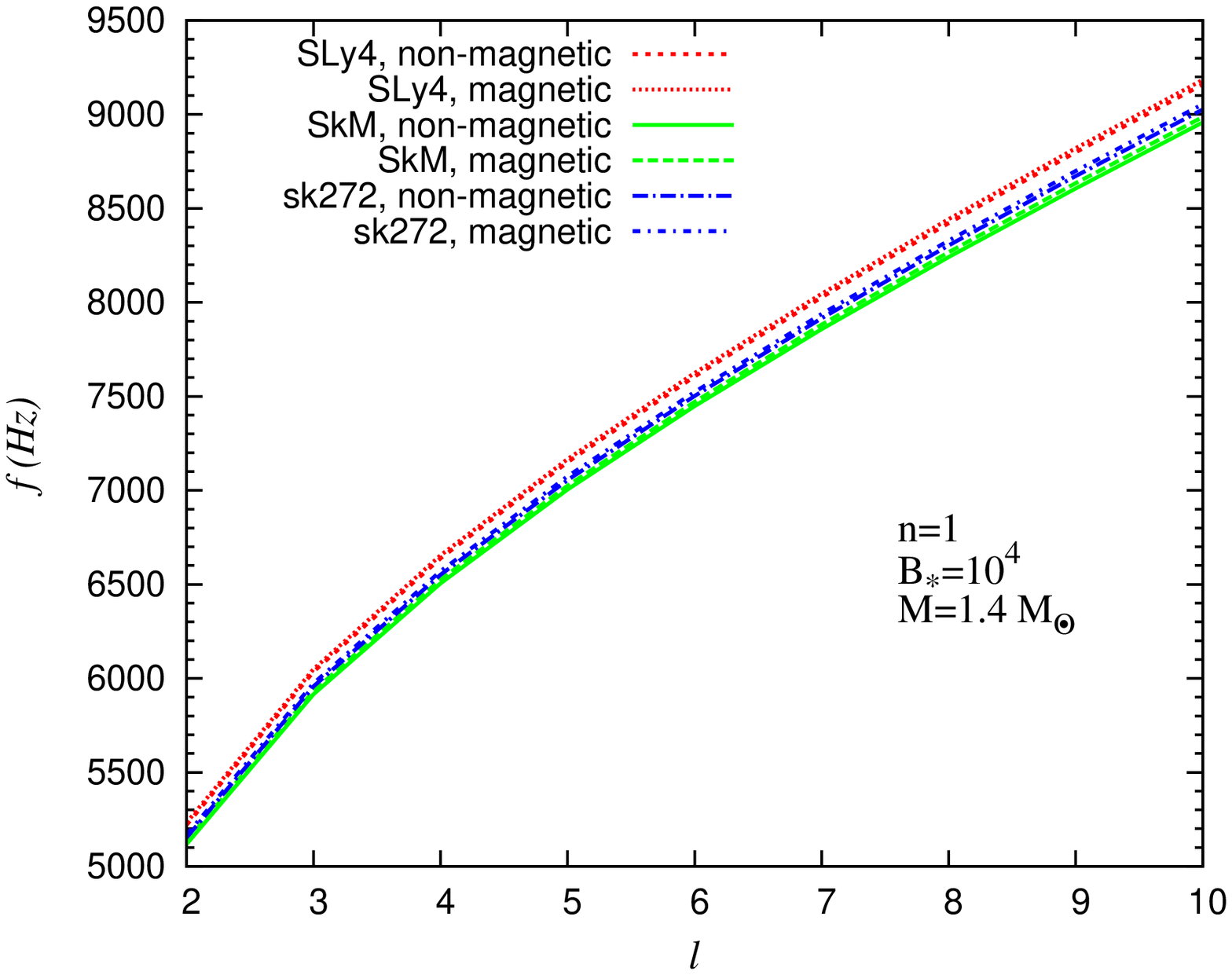}
}}

\vspace{4.0cm}

\noindent{\small{Fig. 8. 
GME mode frequencies for $n=1$ are plotted as a function of 
$\ell$ values 
with and without magnetic crusts of a neutron star of mass $1.4M_{\odot}$ 
based on the SLy4, SkM and Sk272 nucleon-nucleon interactions for 
$B_*=10^4$.}}

\end{document}